\documentclass[fleqn,usenatbib]{mnras}

\usepackage{txfonts}

\usepackage[T1]{fontenc}

\DeclareRobustCommand{\VAN}[3]{#2}
\let\VANthebibliography\thebibliography
\def\thebibliography{\DeclareRobustCommand{\VAN}[3]{##3}\VANthebibliography}

\usepackage{graphicx}			
\usepackage{amssymb}			
\usepackage{threeparttable} 	
\usepackage{etoolbox} 			
\usepackage{booktabs}			
\usepackage{multirow}			
\usepackage{subcaption}			
\usepackage[table,xcdraw]{xcolor}		
\usepackage{multirow}  			
\usepackage{colortbl} 			
\usepackage{tabularx}

\newcommand{\src}{GX~9+9\,}
\graphicspath{{Figures_new/}}

\title[Spectro-polarimetric view of \src]
{Spectro-polarimetric view of bright atoll source \src using \textit{IXPE} and \textit{AstroSat}}

\author[Rwitika et al.]{
Rwitika Chatterjee$^{1}$\thanks{E-mail: rwitika@ursc.gov.in},
Vivek K. Agrawal$^{1}$,
Kiran M. Jayasurya$^{1}$, Tilak Katoch$^{2}$
\\
$^{1}$Space Astronomy Group, ISITE Campus, U. R. Rao Satellite Center, ISRO, Bengaluru 560037, India\\
$^{2}$Department of Astronomy and Astrophysics, Tata Institute of Fundamental Research, Homi Bhabha Road, Colaba, Mumbai 400005, India
}

\date{Accepted XXX. Received YYY; in original form ZZZ}

\pubyear{2023}

\begin{document}
\label{firstpage}
\pagerange{\pageref{firstpage}--\pageref{lastpage}}
\maketitle

\begin{abstract}
We have carried out the first spectro-polarimetric study of the bright NS-LMXB \src using \textit{IXPE} and \textit{AstroSat} observations. We report a significant detection of polarization of $1.7\pm 0.4\%$ over the $2-8$~keV energy band, with a polarization angle of $63^{\circ}\pm 7^{\circ}$. The polarization is found to be energy-dependent, with a $3\sigma$ polarization degree consistent with null polarization in $2-4$~keV, and $3.2\%$ in $4-8$~keV. Typical of the spectra seen in NS-LMXBs, we find that a combination of soft thermal emission from the accretion disc and Comptonized component from the optically thick corona produces a good fit to the spectra. We also attempt to infer the individual polarization of these components, and obtain a $3\sigma$ upper limit of $\sim 11\%$ on the polarization degree of the thermal component, and constrain that of the Comptonized component to $\sim 3\%$. We comment on the possible corona geometry of the system based on our results.
\end{abstract}

\begin{keywords}
accretion, accretion discs - polarization - X-rays: binaries - X-rays: individual: GX 9+9 
\end{keywords}



\section{Introduction}

Atoll and Z-sources are luminous low-mass X-ray binaries (LMXBs) where a neutron star accretes matter from a low-mass companion through Roche-lobe overflow. These two types of LMXBs trace different patterns on their colour-colour diagram (CCD, \citealt{hasinger1989}). Atoll sources exhibit high variability over timescales ranging from milliseconds to years, and depending on their X-ray luminosity, may be found either in the high soft state (HSS) or low hard state (LHS).
\par X-ray spectra of the NS-LMXBs are complex and require multiple components to describe them.  A combination of multi-temperature disc blackbody and Comptonized emission (\citealt{disalvo2000a, disalvo2002, agrawal2003, tarana2008, agrawal2009, agrawal2023}) has been widely used to describe the X-ray spectra of the atoll and Z-sources. Here, the softer component from the disc is directly observed, and the Comptonized emission may come from the boundary layer (BL) or spreading layer (SL) near the surface of the neutron star (`eastern model'). An alternative scenario is that the accretion disc is Comptonized and blackbody emission comes from the neutron star surface (`western model'), and a combination of blackbody plus a Comptonized component is adopted for describing the spectra of these sources (\citealt{piraino2000, piraino2007, disalvo2000b, disalvo2001, barret2002, wang2019}). 
\par \src is an NS-LMXBs, classified as an atoll source. It was discovered by a sounding rocket experiment on 1967~July~7 \citep{bradt1968}. \citet{hertz1988} reported a $4.19\pm 0.02$~h orbital period using scanning mode of \textit{HEAO~A-1}. Several models have been used to describe the spectrum of this source, such as power-law with exponential cutoff \citep{church2001}, blackbody plus Comptonization model \citep{kong2006}, and multicolor disc (MCD) plus Comptonized blackbody emission from the SL \citep{savolainen2009}. \citet{gogus2007} described the \textit{RXTE-PCA} spectrum of this source by a double blackbody plus a broad iron line at 6~keV. \citet{iaria2020} reported the presence of a blurred reflection component using $0.3-40$~keV spectrum of the source.
\par Spectral modeling of such sources helps to understand the nature of the coronal plasma (e.g. temperature, optical depth). However, these models are degenerate with respect to the geometry and location of corona, and polarization information, if available, can be used to constrain them. Recent polarimetric studies of several NS-LMXBs such as Sco~X-1 \citep{long2022}, Cyg~X-2 \citep{farinelli2023} and GS~$1826-238$ \citep{capitanio2022} have revealed the capability of polarization studies in inferring the source geometry and emission properties. In this letter, we present the first spectro-polarimetric study of the bright atoll source \src using \textit{IXPE} observations and archival \textit{AstroSat} data.    

\section{Observations and Data Analysis}
\label{sec:obs_data}
\subsection{\textit{AstroSat}}
The source GX 9+9 was observed from 2020~July~25 to 2020~July~27 with {\it AstroSat} for a net exposure time of 56~ks. To understand the spectral nature of the source, we analyzed data from the Soft X-ray Telescope ({\it SXT}, $0.3-8$~keV) and Large Area X-ray Proportional Counter ({\it LAXPC}, $3-60$~keV), which were operated in photon-counting mode and event analysis mode respectively.
\par HEASOFT FTOOL\footnote{http://heasarc.gsfc.nasa.gov/ftools} \texttt{XSELECT} was used to create the image, spectra and light curves from the Level-2 data of {\it SXT}. The {\it SXT} count rate during the observation was above the pileup limit. Hence, we used an annular region excluding the central 2$\arcmin$, and with outer radius 12$\arcmin$ to create light curves and spectra. ARF file was created using the task \texttt{sxtARFModule}\footnote{http://www.tifr.res.in/$\sim$astrosat\_sxt/dataanalysis.html}. \textit{LAXPC} light curves and spectra were generated by the latest version of the {\it LAXPC}  analysis software \texttt{LaxpcSoft}\footnote{http://www.tifr.res.in/$\sim$astrosat\_laxpc/LaxpcSoft.html}. The background spectra of the dark sky was provided by the instrument team and was used for the analysis. 

\begin{figure}
\centering
\includegraphics[width=0.4\textwidth,angle=-90, trim = 0cm 1.4cm 0cm 0cm, clip=True]{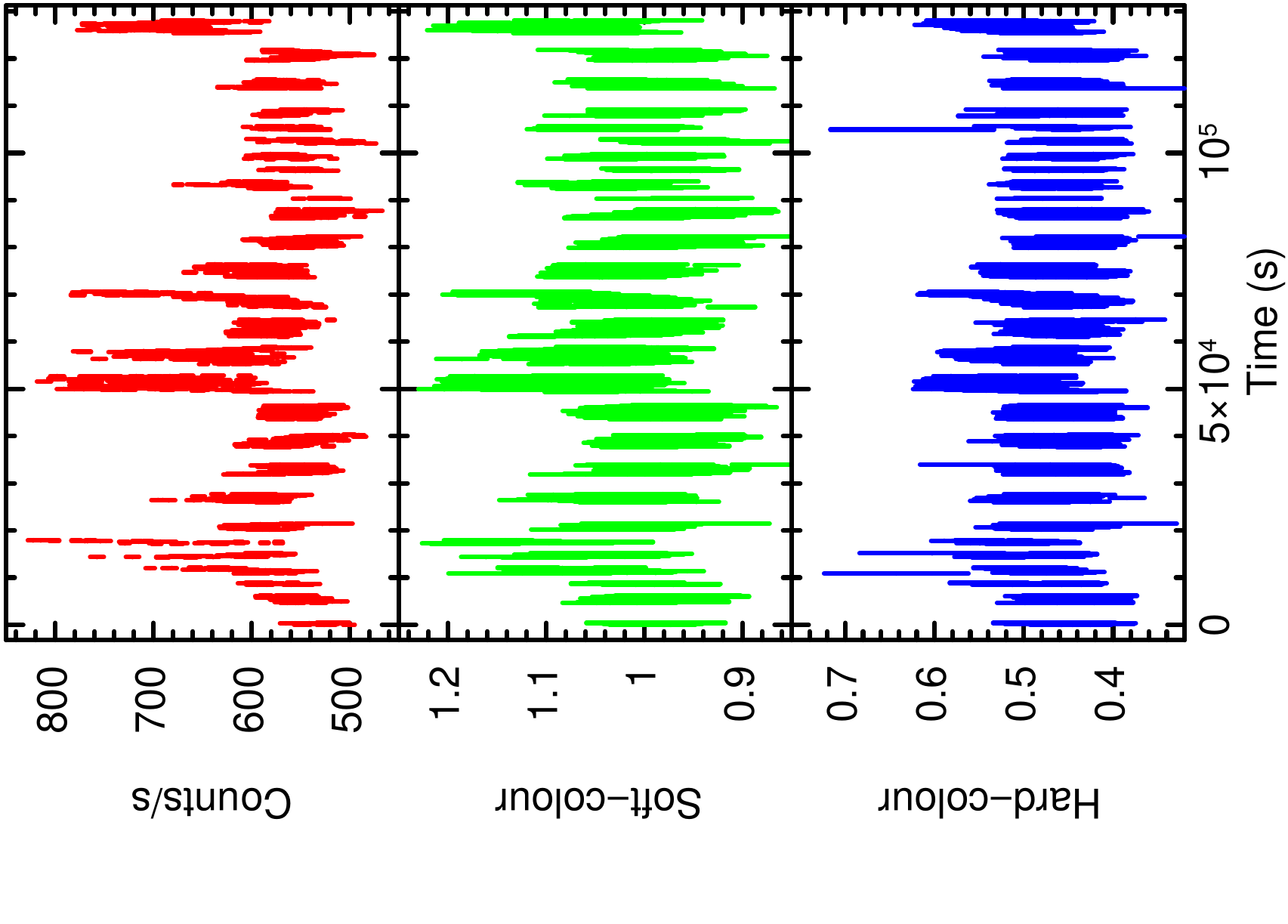}
\caption{\textit{LAXPC} $3-25$~keV count rate (top), soft-colour (middle) and hard-colour (bottom) during the \textit{AstroSat} observation. Bin time of 16~s has been used to generate the light curve and hardness plots.} 
\label{fig:lc}
\end{figure}

\begin{figure}
\centering
\includegraphics[width=0.42\textwidth, trim = 0cm 0cm 0cm 1.5cm, clip=True]{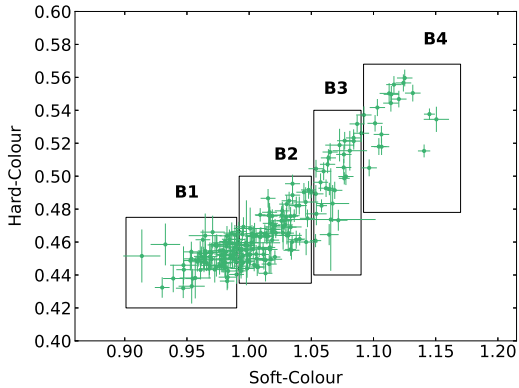}
\caption{CCD of \src, with different parts denoted by labels B1, B2, B3 and B4. Each data points corresponds to 256~s of integration time.}
\label{fig:ccd}
\end{figure}

\par Figure~\ref{fig:lc} shows the variation of the \textit{LAXPC} $3-25$~keV count rate, soft-color and hard-colour as a function of time, over the \textit{AstroSat} observation. We defined soft-colour as the ratio of count rates in $5-7$~keV and $3-5$~keV, and hard colour as the ratio of count rates in $10-20$~keV and $7-10$~keV energy bands. We also produced the CCD of the source, which is shown in Figure~\ref{fig:ccd}. We note that the source is in the `banana' state during the observation. We divided the CCD into 4 sections and denote them by B1, B2, B3 and B4.
\par We created spectra for the four sections and performed joint spectral fitting of each section using the $0.7-7$~keV spectra from \textit{SXT} and $4-25$~keV spectra from \textit{LAXPC}\footnote{Beyond 25~keV, the source is weak and the background dominates.} using \textit{XSPEC}\footnote{https://heasarc.gsfc.nasa.gov/xanadu/xspec/} \citep{arnaud1996} v12.12.1. We found that a Comptonized emission component (\citealt{zdziarski1996} \texttt{nthcomp} in \textit{XSPEC}) plus a thermal component, either described by a single temperature blackbody or an MCD (\texttt{bbodyrad} and \texttt{diskbb} respectively), modified by ISM absorption (\texttt{tbabs}), give a satisfactory fit to the $0.7-25$~keV spectra of the source\footnote{We also need two edge components at 2.4~keV and 8.7~keV (instrumental features) to obtain a good fit.}. The seed photons for Comptonization are assumed to follow a blackbody distribution (\texttt{inp\_type} parameter set to 0 in \texttt{nthcomp}), when the thermal component comes from the accretion disc (hereafter, model~1). However, multi-temperature disc blackbody is used for seed photons (\texttt{inp\_type} = 1) when thermal component is described by \texttt{bbodyrad} (hereafter, model~2). We note here that model~1 is reminiscent of the `eastern' type model where thermal component from the disc is directly observed and the Comptonized emission comes from a shell-like corona around the neutron star/SL, whereas model~2 resembles a `western'-like scenario, with the corona as a slab or wedge above the accretion disc. 

\subsection{\textit{IXPE}}
\begin{figure}
\centering
\includegraphics[scale=0.32, trim = 0.7cm 0cm 0cm 0cm, clip=True]{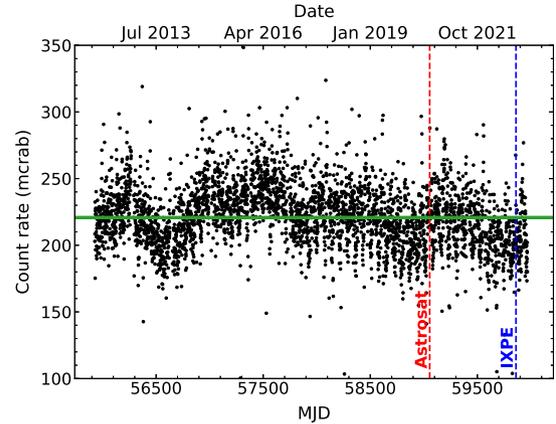}
\caption{\textit{MAXI/GSC} $2-20$~keV light curve of \src from 2012 till present day. The epochs of the \textit{AstroSat} and \textit{IXPE} observations are marked. The horizontal line indicates the mean count rate of 220~mcrab.}
\label{fig:maxi}
\end{figure}
\par The Imaging X-ray Polarimetry Explorer (\textit{IXPE}, \citealt{weisskopf2022}), launched on 2021 December 9, consists of three identical telescopes each consisting of a mirror module assembly with a polarization-sensitive imaging X-ray detector at the focus. \textit{IXPE} observed GX~9+9 on 2022 October 9 for about 92~ks of net exposure time. In Figure~\ref{fig:maxi}, we show the long-term \textit{MAXI/GSC} light curve of the source and note that the source is stable and remains in the HSS over the period of both \textit{AstroSat} and \textit{IXPE} observations.
\par We analyzed the processed Level-2 data\footnote{Publicly available on the HEASARC archive} using \texttt{IXPEOBSSIM} software v30.0.0 \citep{baldini2022}. Source was defined as a circular region of radius 60" centered at the intensity centroid, and background as an annular region with the same center and inner and outer radii as 180" and 240" respectively. Source and background events were filtered using \texttt{XPSELECT} task and various binning algorithms were applied using \texttt{XPBIN} task. The average count rate was 13 cts/s with the background contribution $\lesssim 0.2\%$. The polarization parameters (polarization degree PD and polarization angle PA) were extracted in the energy bands $2-4$~keV, $4-8$~keV and $2-8$~keV using the model-independent PCUBE algorithm \citep{kislat2015}. For calculating the uncertainties, PCUBE assumes that the PD and PA are independent whereas they are not actually so. Hence, the contours of the joint measurement of PD and PA represent the uncertainties more appropriately. 
\par We also performed a spectro-polarimetric model-dependent fit (see e.g. \citealt{strohmayer2017}) of the data using \textit{XSPEC}. Source and background spectra corresponding to the Stokes parameters I, Q and U were extracted using algorithms PHA1, PHA1Q and PHA1U respectively, and the latest response files (v12) were used in spectral fitting. The best-fit models obtained from the joint \textit{SXT} and \textit{LAXPC} fits (see Section~\ref{sec:spec_res}), modified by a constant polarization (\texttt{polconst}\footnote{Assumes a constant PD and PA over the energy range of interest.} in \textit{XSPEC}), was used in the fitting.   

\section{Results}
\subsection{Spectral properties}
\label{sec:spec_res}

\begin{table}
\centering
	\caption{Best-fit parameters of model~1 to \textit{SXT+LAXPC} spectra.  $N_H$: hydrogen column density (10$^{22}$ cm$^{-2}$); $\Gamma$: photon index; $\tau$: optical depth; $kT_{in}$: inner disc temperature (keV); $kT_e$: plasma temperature (keV). $N_{dbb}$ and $N_{nth}$ are the normalizations of the \texttt{diskbb} and \texttt{nthcomp} components. 90\% uncertainties are quoted.}
	\label{tab:spec}
	\begin{threeparttable}
	\bgroup
	\setlength{\tabcolsep}{2.3pt}
\def\arraystretch{1.4}
{
	\begin{tabular}{cccccc}
\hline
Parameters & B1 & B2 & B3 & B4 & Average\\
\hline
$N_H$      &0.17$\pm$0.01 & 0.16$\pm$0.01 & 0.15$\pm$0.01 &0.16$\pm$0.01 & 0.17$\pm$0.01 \\
$kT_{in}$ & 0.68$\pm$0.05  &0.65$\pm$0.08 &0.77$\pm$0.09 &0.96$\pm$0.06 & 0.65$\pm$0.04\\
$N_{dbb}$         &345$\pm$75 &373$\pm$127 & 221$\pm$62 & 108$\pm$18 & 405$\pm$64\\
$\Gamma$        &3.46$\pm$0.24 & 3.10$\pm$0.19 & 3.11$\pm$0.24 & 2.62$\pm$0.09 & 3.14$\pm$0.14 \\
$kT_e$          &7.55$\pm$1.78 & 5.12$\pm$0.59 & 5.56$\pm$1.51 & 4.20$\pm$0.35 & 5.58$\pm$0.82\\
Seed $kT$\tnote{a}   & 1.05$\pm$0.05 & 1.03$\pm$0.06 & 1.11$\pm$0.06 & 1.24$\pm$0.13 &1.05$\pm$0.03\\
$N_{nth}$        & 0.15$\pm$0.02  & 0.16$\pm$0.02 & 0.12$\pm$0.02 & 0.10$\pm$0.01 & 0.15$\pm$0.01\\
$\tau$\tnote{b}        &2.88$\pm$0.72 & 3.92$\pm$0.48 & 3.76$\pm$0.77 & 5.5$\pm$0.47 & 3.72$\pm$0.69\\
$\chi^2$/DOF & 598/654 & 656/654 & 608/654 & 683/654 & 628/654\\
\hline

\end{tabular}
}
\egroup
\begin{tablenotes}
	\item[a] Blackbody seed photons
	\item[b] Assuming spherical corona
	\end{tablenotes}
	\end{threeparttable}
\end{table}

We show the best fit spectral parameters obtained by fitting the different sections of the atoll track using model~1 in Table~\ref{tab:spec}. As the source moves along the banana track from B1 to B4, the spectrum becomes harder, with the photon index decreasing from 3.5 to 2.6. The temperature at the inner radius of the accretion disc increases from 0.7~keV to $\sim 1$~keV, and the seed photon temperature ranges from 1~keV to 1.2~keV. The temperature of the corona varies in the range 4.2~keV to 7.6~keV. The corona is optically thick, with $\tau$ in the range $\sim 2.9-5.5$. 

\begin{figure}
\centering
\includegraphics[width=0.28\textwidth,angle=-90, trim = 0cm 1.4cm 0cm 0.cm, clip=True]{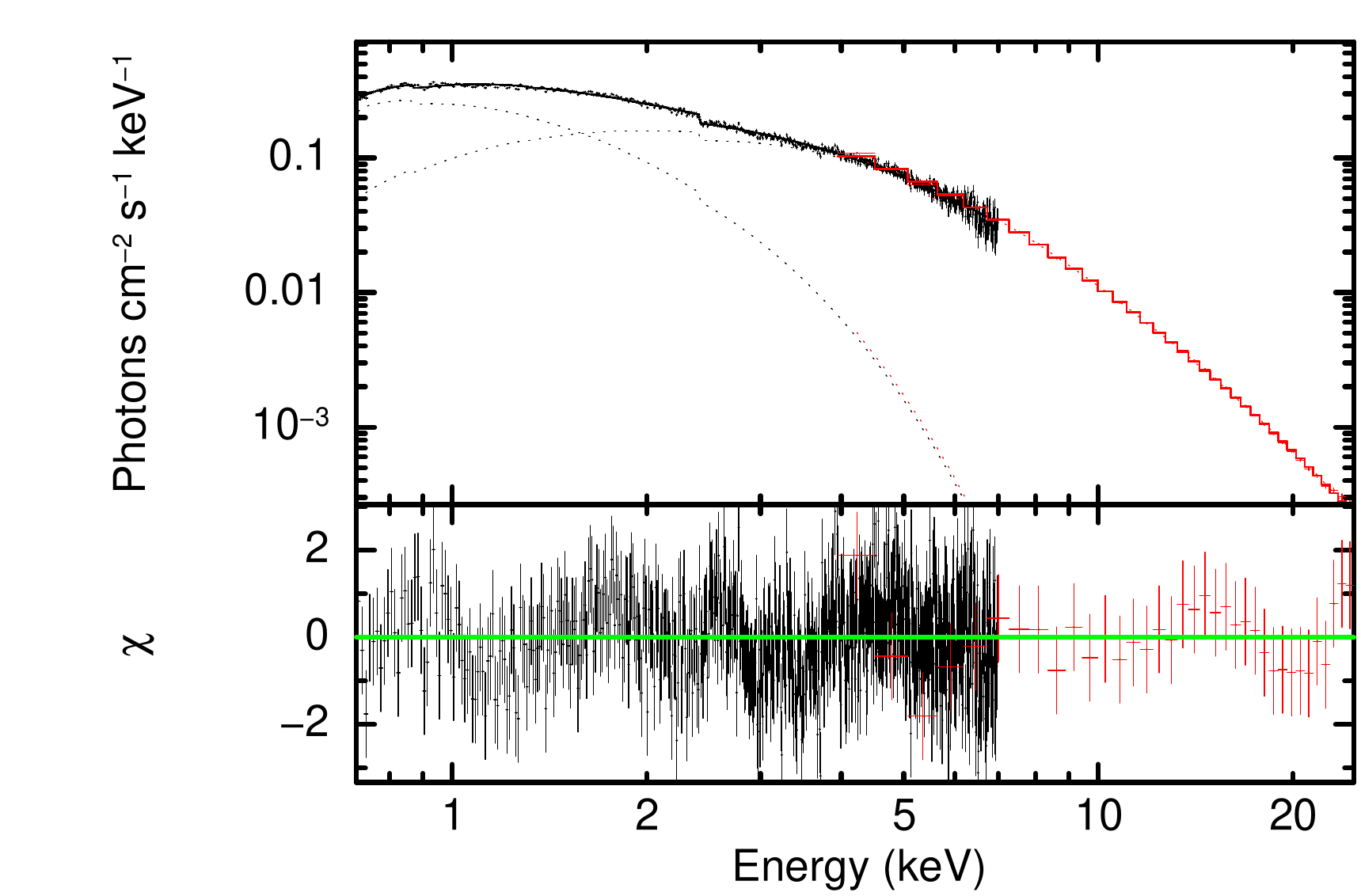}
	\caption{(Top) Unfolded average X-ray spectrum (\textit{SXT}:black, \textit{LAXPC}: red) of \src, fitted with model~1. (Bottom) Residuals in units of $\sigma$.}
	\label{fig:bestfit}
\end{figure}
\par It is evident from Figures~\ref{fig:lc} and \ref{fig:ccd} that soft-colour and hard-colour vary significantly and can trace a significant portion of the banana track on the timescale of a few hours. We also note that the spectral properties show only subtle variations as the source moves along the CCD. The PD and PA have been calculated using $\sim 90$~ks of \textit{IXPE} observations (Section~\ref{sec:pol_res}) and it is difficult to identify the location of the source on the CCD during the observations. Hence, we also constructed an average spectrum across all the sections of the CCD (B1 to B4), which can be described by a combination of disc blackbody with $kT_{in}\sim 0.6$~keV and $5.6$~keV corona (see Table~\ref{tab:spec} and Figure~\ref{fig:bestfit}). We note that model~2 is able to fit the spectra equally well, and the average spectrum can also, equivalently, be described by a 1.4~keV blackbody and 0.8~keV disc (seed) photons Comptonized by a 3.7~keV corona.

\subsection{Spectro-polarimetric properties}
\label{sec:pol_res}
\begin{table}
\centering
\caption{Polarimetric parameters obtained from PCUBE in different energy bands. Values reported are for all three DUs combined. Associated $1\sigma$ uncertainties are quoted.}
\label{tab:pcube}
\bgroup
\def\arraystretch{1.4}
{
\begin{tabular}{cccc}
\hline
Parameter               & $2-4$ keV                 & $4-8$ keV  & $2-8$ keV  \\ \hline
Q/I (\%)                & $-0.75\pm 0.38$				 & $-1.36\pm 0.72$ & $-0.96\pm 0.39$ \\
U/I (\%)                & $0.57\pm 0.38$                 & $2.89\pm 0.72$  & $1.36\pm 0.39$  \\
PD (\%)                 & $0.94\pm 0.38$                 & $3.19\pm 0.72$  & $1.66\pm 0.39$  \\
PA (deg)         & $71.4\pm 11.5$                 & $57.6\pm 6.4$   & $62.6\pm 6.7$   \\ \hline
\end{tabular}%
}
\egroup
\end{table}

\begin{figure*}
\centering
\includegraphics[scale=0.4, trim = 3cm 0cm 2cm 0cm, clip=True]{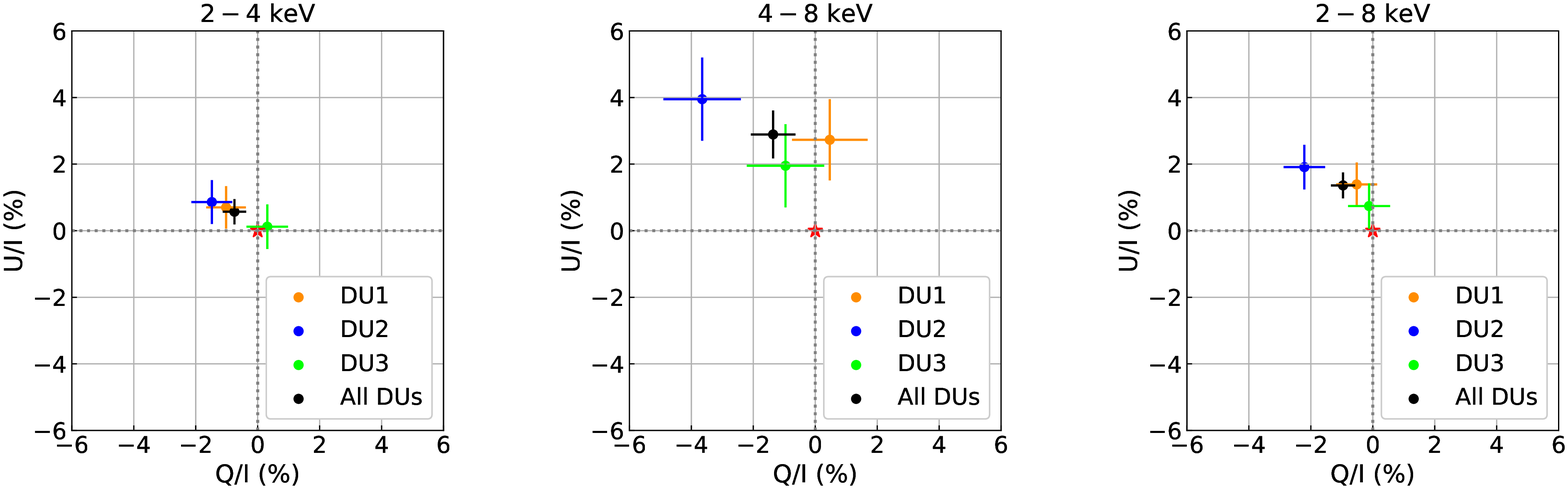}
\caption{Normalised stokes parameters (Q/I and U/I) obtained using PCUBE in the energy bands $2-4$~keV (left), $4-8$~keV (middle) and $2-8$~keV (right), for DU1 (orange), DU2 (blue), DU3 (green) and for the three DUs combined (black). In each panel, the red star corresponds to null polarization.}
\label{fig:stokes}
\end{figure*}

The results obtained using PCUBE for the three identical detector units (DUs) combined, in $2-4$~keV, $4-8$~keV and $2-8$~keV energy bands, are summarized in Table~\ref{tab:pcube}. The normalized stokes parameters of each DU and all three DUs combined are also presented graphically in Figure~\ref{fig:stokes}. In $2-4$~keV, the measured PD is below the minimum detectable polarization (MDP) at the $99\%$ level. However, in $4-8$~keV as well as over the entire \textit{IXPE} energy range of $2-8$~keV, we report a significant detection of polarization ($4.4\sigma$ and $4.2\sigma$ respectively), with PD = 3.2\% and 1.7\%, and PA = $57.6^{\circ}\pm 6.4^{\circ}$ and $62.6^{\circ}\pm 6.7^{\circ}$ respectively. 

\begin{table}
\centering
\caption{Spectro-polarimetric parameters obtained by simultaneously fitting Stokes spectra in \textit{XSPEC} in different energy bands using model~1. Results using single \texttt{polconst} as well as two \texttt{polconst} (one for each spectral component) are reported. Uncertainties, upper limits quoted are at $3\sigma$ level.}
\label{tab:polxspec}
\begin{threeparttable}
\bgroup
\setlength{\tabcolsep}{4pt}
\def\arraystretch{1.4}
{
\begin{tabular}{cccccc}
\hline
\multirow{2}{*}{Parameter}     & \multicolumn{3}{c}{Single \texttt{polconst}} & \multicolumn{2}{c}{Multiple \texttt{polconst}\tnote{*}}  \\\cline{2-6}
                           & $2-4$ keV           & $4-8$ keV          & $2-8$ keV          & \texttt{diskbb}     & \texttt{nthcomp}   \\ \hline
PD (\%)                    & $< 2.03$      & $2.77\pm 1.92$     & $1.38\pm 0.99$     & $< 10.98$                       & $2.96\pm 2.01$       \\
PA ($^{\circ}$)            & ...      & $62.1\pm 21.9$      & $68.4\pm 23.3$      & ...                      & $66.4\pm 21.4$                  \\
$\chi^2$/DOF               & 400/434           & 839/884          & 1281/1334        & \multicolumn{2}{c}{1274/1332}                             \\ \hline
\end{tabular}%
}
\egroup
\begin{tablenotes}
	\item[*] Fitted over $2-8$~keV
	\end{tablenotes}
	\end{threeparttable}
\end{table}

\begin{figure*}
\centering
\begin{subfigure}{0.6\columnwidth}
\includegraphics[scale=0.25, trim = 0cm 0cm 6.5cm 0cm, clip=true]{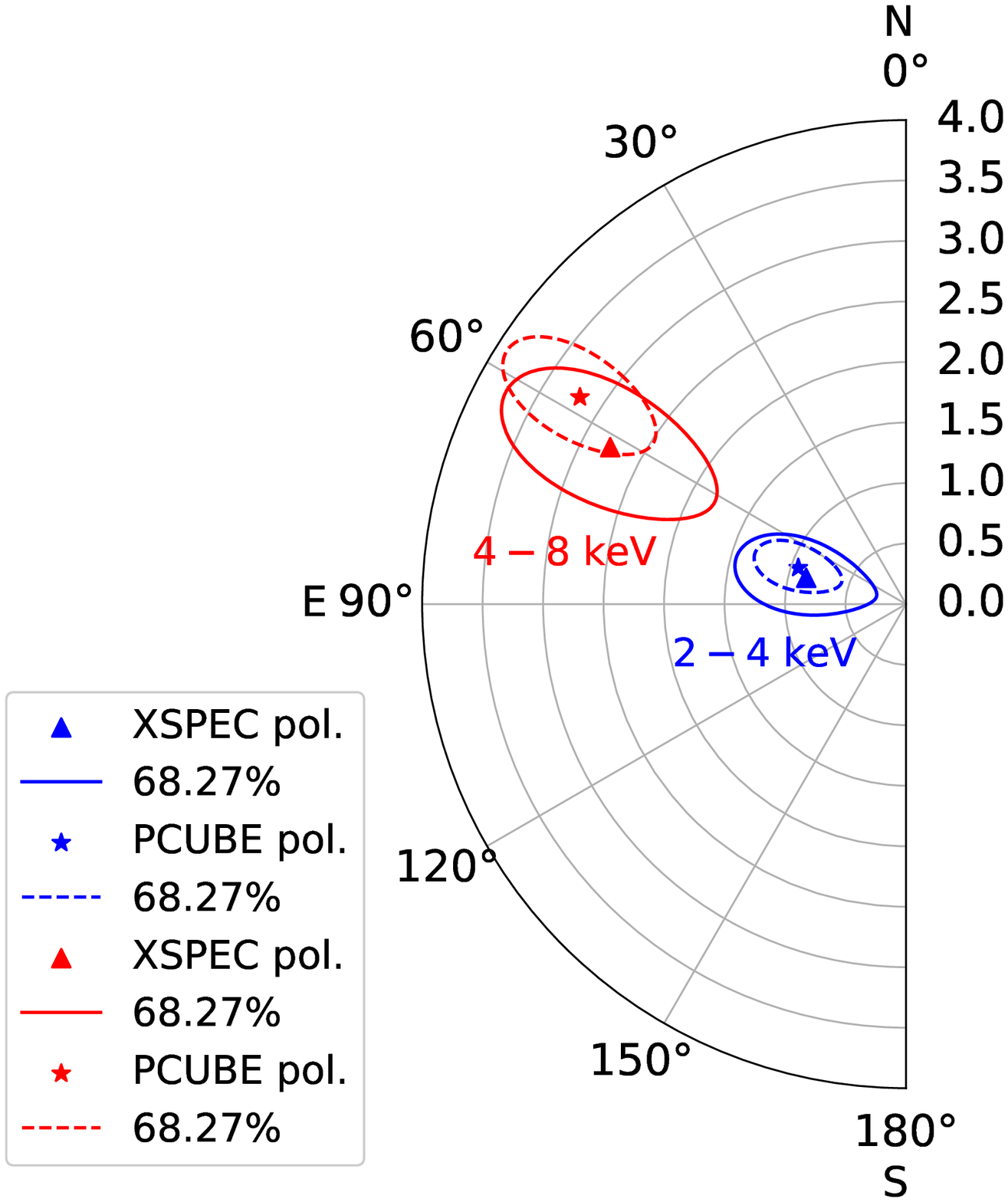}
\caption{}
\end{subfigure}\hfill
\begin{subfigure}{0.6\columnwidth}
\includegraphics[scale=0.25, trim = 0cm 0cm 5.6cm 0cm, clip=true]{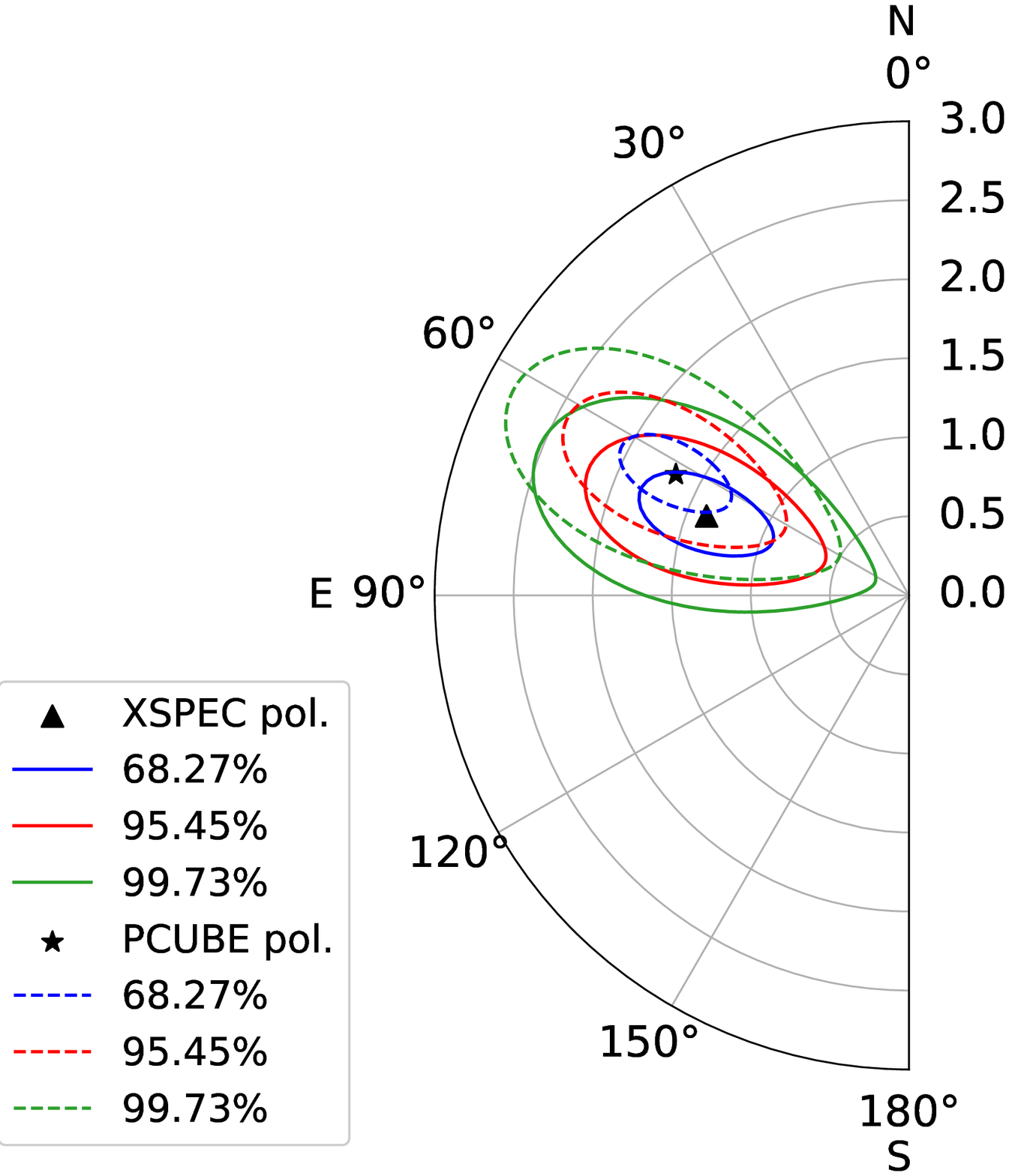}
\caption{}
\end{subfigure}\hfill
\begin{subfigure}{0.6\columnwidth}
\includegraphics[scale=0.25, trim = 1.5cm 0cm 6cm 0cm, clip=true]{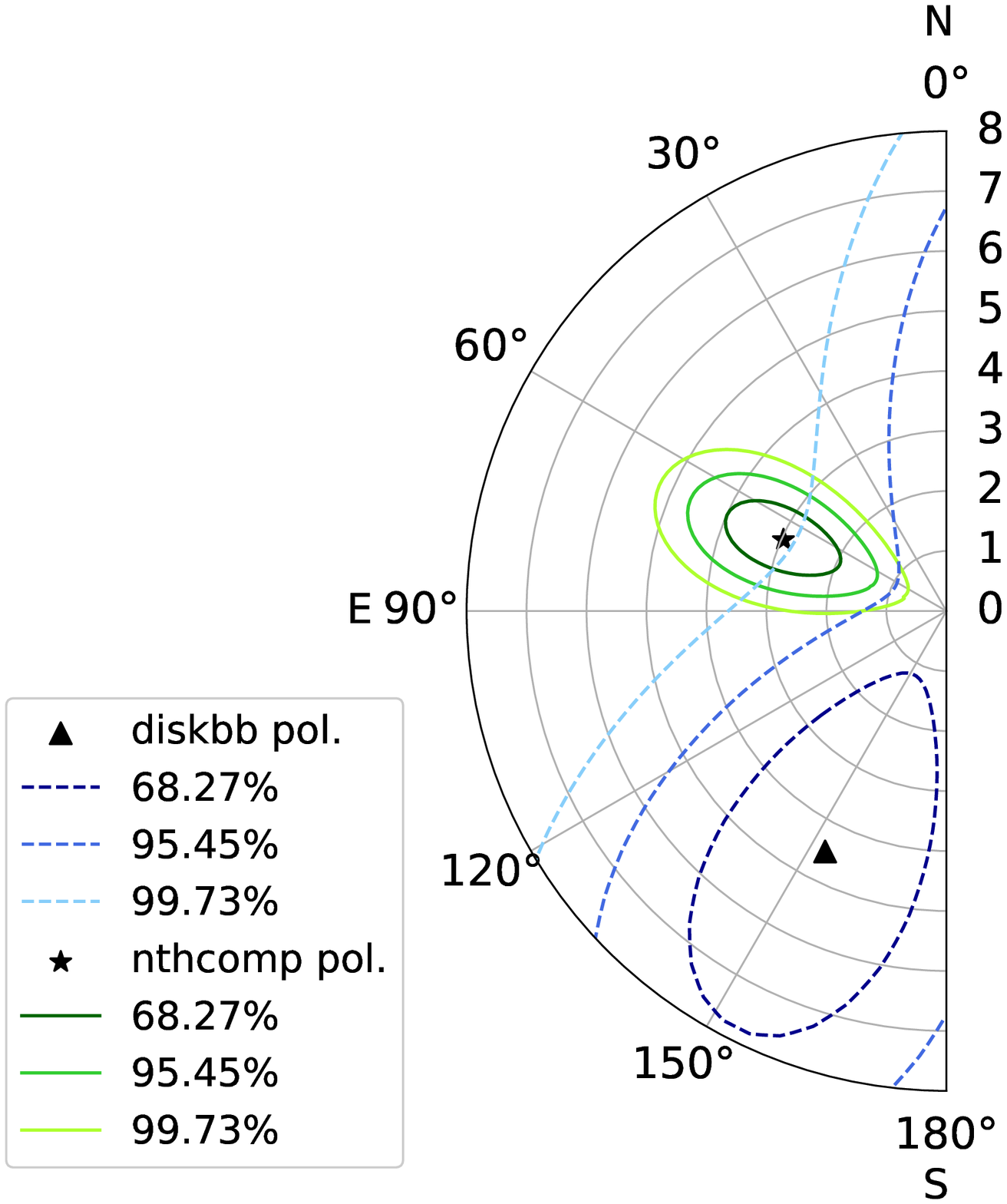}
\caption{}
\end{subfigure}\hfill
\caption{(a) Results of PCUBE (star, dashed lines) and \textit{XSPEC} (triangle, solid lines) in the $2-4$~keV (blue) and $4-8$~keV (red) energy bands. $1\sigma$ contours are plotted (b) Results of PCUBE (star, dashed lines) and \textit{XSPEC} (triangle, solid lines) over $2-8$~keV. $1\sigma$ (blue), $2\sigma$ (red) and $3\sigma$ (green) contours are shown. (c) Individual \texttt{polconst} for \texttt{diskbb} (triangle, blue contours) and \texttt{nthcomp} (star, green contours). In all three panels, the \textit{IXPE} spectrum has been fitted with the parameters of the average spectrum from \textit{AstroSat}, fitted with model~1. See text for details.}
\label{fig:pol_contour}
\end{figure*} 

\par To verify the robustness of the polarization parameters obtained with PCUBE, we also followed a model-dependent spectro-polarimetric approach. We simultaneously fitted the Stokes I, Q and U spectra of all three DUs with model~1 (Section~\ref{sec:spec_res}), multiplied by \texttt{polconst}. Owing to statistics and limited energy band of \textit{IXPE}, the parameters $N_H$, $kT_{in}$, $\Gamma$ and $kT_e$ were frozen to the best fit values obtained by fitting the average spectrum. We performed the spectral fits for the three energy bands as defined earlier, and the results are shown in Table~\ref{tab:polxspec} and Figures~\ref{fig:pol_contour}a and \ref{fig:pol_contour}b. The obtained PD and PA values are consistent with those obtained via PCUBE, within uncertainties. The contours, although not entirely coincident, are in good agreement with each other.  
\par Motivated by the energy-dependence of the PD, we attempted to compute the PD corresponding to each spectral component, and hence we fitted the model \texttt{tbabs*(polconst*diskbb+polconst*nthcomp)} to the spectra. The results are shown in Table~\ref{tab:polxspec} and Figure~\ref{fig:pol_contour}c. The polarization of \texttt{diskbb} could be constrained at only the $1\sigma$ level, with PD = 4.49\% and PA = 153.2$^{\circ}$. However, the polarization of \texttt{nthcomp} is well-constrained, with PD = 2.96\% and PA = 66.4$^{\circ}$, roughly perpendicular to that of \texttt{diskbb}.
\par We also repeated the spectro-polarimetric fitting using model~2, with the spectral parameters frozen to the best-fit values corresponding to the average spectrum. Using a single \texttt{polconst} to describe the emission, we obtained equivalent fits as with model~1, with similar PD and PA values in $2-4$~keV, $4-8$~keV and $2-8$~keV energy bands. However, we were not able to constrain the polarization properties of the individual model components (\texttt{bbodyrad} and \texttt{nthcomp}) at the $3\sigma$ level.

\section{Discussion}
\label{sec:discuss}
In this letter, we attempt to understand the geometry of the X-ray emitting region and radiative processes in the bright atoll X-ray binary \src using {\it IXPE} and {\it AstroSat} data. As in the case of several NS-LMXBs, it has been observed that X-ray spectral fits of this source with respect to different models are degenerate and produce equivalent fits to its spectrum. In particular, it is of interest to determine the location and geometry of the Comptonizing corona, and polarization information along with spectroscopy can potentially distinguish between the models.
\par In the present work, we found that the spectrum of \src can be described by an optically thick Comptonized emission plus a thermal component. The thermal component can be represented by either a MCD (model~1) or blackbody (model~2), and both these approaches produce statistically good descriptions of the data. To break the spectral degeneracy in the NS-LMXBs, sensitive polarization measurements were long-awaited and thanks to the superior quality of polarimetric data from \textit{IXPE}, it is now possible to attempt to distinguish between the different spectral models, and also provide an important probe of the geometry of the corona (see e.g. \citealt{farinelli2023}).
\par An analysis of the \textit{IXPE} data of \src reveals that the $2-8$~keV X-ray emission of the source is polarized, with PD = 1.7\% and PA $\sim 63^{\circ}$. The polarization is higher in the $4-8$~keV (3.2\% at $4.4\sigma$) band compared to that in the $2-4$~keV band (0.9\% at $2.5\sigma$). Recently, \citet{gnarini2022} carried out detailed polarimetric simulations of NS-LMXBs considering two different coronal geometries, viz. shell and slab. The authors reported the expected PD and PA from such sources considering different system inclinations and spectral states (HSS or LHS). The inclination of the \src system lies between $40^{\circ}-60^\circ$ (\citealt{iaria2020,hertz1988}) and the source was found to be in the HSS (Figures~\ref{fig:ccd} and \ref{fig:maxi}). The simulations suggest that at these source inclinations, the PA in $2-8$~keV is less than 90$^\circ$ for shell geometry and always stays greater than 100$^\circ$ for slab geometry when the source is in HSS. Our estimate of PA $\sim63^\circ$ supports the shell type corona geometry. The spherical shell could be in the form of a SL above the neutron star surface. However, we also note here that the predicted trend for shell geometry (PD decreasing with increasing energy, see \citealt{gnarini2022}) does not agree with the present results. 
\par A similar shell-type geometry was also proposed by \citet{farinelli2023} for Cyg~X-2 using polarization measurements from \textit{IXPE}, wherein the neutron star surface provides the seed photons and the surface is covered by a spherical corona or spreading layer. The uncovered disc emits the thermal component. Motivated by this, we carried out model-dependent studies by fitting the \textit{IXPE} data with two spectral models generally associated with these two corona geometries (shell: model~1, slab: model~2). Both these models give similar PD and PA as obtained from PCUBE. Using model~1, we were also able to ascertain the polarization of the individual spectral components and the results suggest that the Comptonized component is polarized and the disc emission is possibly unpolarized. \citet{farinelli2023} obtained a similar result for Cyg X-2. 
\par In the case of Sco~X-1 and Cyg~X-2, the PA was found to coincide with the system symmetry axis (\citealt{long2022, farinelli2023}). Unfortunately, \src is radio faint \citep{vandeneijnden2021} and a jet has not yet been detected from the source. Provided the polarization is along the system symmetry axis, which is perpendicular to the accretion disc, our results may indicate that the Comptonized emission originates in the boundary layer or transition layer, as in the case of Sco~X-1. Hence, simulations for alternative geometries such as transition layer or wedge-like corona can provide further insights on the polarization data from bright NS-LMXBs such as \src. In addition, more polarimetric observations of LMXBs are required to address the geometry of the accretion flow for different classes of NS-LMXBs in different spectral states.

\section*{Acknowledgments}
RC, VKA and KMJ thank GH, SAG; DD, PDMSA and Director, URSC for encouragement and continuous support to carry out this research. We thank Prof. Rosario Iaria for his comments which helped to improve the quality of the manuscript.
\par This research used data products provided by the \textit{IXPE} Team (MSFC, SSDC, INAF, and INFN) and distributed with additional software tools by the High-Energy Astrophysics Science Archive Research Center (HEASARC), at NASA Goddard Space Flight Center (GSFC). This work has used the data from the \textit{LAXPC} Instruments developed at TIFR, Mumbai and the \textit{LAXPC} POC at TIFR is thanked for verifying and releasing the data via the ISSDC data archive. This work has used the data from {\it SXT} developed at TIFR, Mumbai, and the {\it SXT} POC at TIFR is thanked for verifying \& releasing the data and providing the necessary software tools.\\
\section*{Data Availability}
Data underlying this work are available at High Energy Astrophysics Science Archive Research Center (HEASARC) facility, located at NASA-Goddard Space Flight Center and \textit{AstroSat}-ISSDC website (http://astrobrowse.issdc.gov.in/astro\_archive/archive). The MAXI light curve used in this work is publicly available at http://maxi.riken.jp/top/slist.html.



\bibliographystyle{mnras}
\bibliography{References}



\bsp	
\label{lastpage}
\end{document}